\documentstyle[aps,prb,twocolumn,graphics]{revtex}

\begin{document}

\twocolumn[\hsize\textwidth\columnwidth\hsize\csname
@twocolumnfalse\endcsname

\title{Inter-layer Hall effect in double quantum wells subject to
 in-plane magnetic fields}

\draft

\author{J. Koloren\v{c}, L. Smr\v{c}ka and P. St\v{r}eda}
\address{Institute of Physics, Academy of Sciences of the Czech
Republic, Cukrovarnick\'{a} 10, 162 53 Praha 6}
\date{\today}
\maketitle

\begin{abstract}
We report on a theoretical study of the transport properties of two
coupled two-dimensional electron systems subject to in-plane magnetic
fields. The charge redistribution in double wells induced by the
Lorenz force in crossed electric and magnetic fields has been
studied. We have found that the redistribution of the charge and the
related inter-layer Hall effect originate in the chirality of
diamagnetic currents and give a substantial contribution to the
conductivity.
\end{abstract}


\vskip2pc]
\section{Introduction}\label{intro}
Semiconductor quantum wells, multi-wells and superlattices are
designed to have an artificial electronic structure determined by
their construction rather than by the properties of individual
semiconductor materials~\,\cite{kelly}. The band profiles of
multi-wells are formed by a sequence of quantum wells separated by
barriers. Electrons condense to two-dimensional (2D) electron layers
localized in the wells and the tunnel-coupling of the layers leads to
the formation of new electron systems, rich in optical and transport
effects, that can be modified by application of the magnetic field.

Two experimental arrangements of the magnetotransport measurement in
multi-well structures subjected to the in-plane magnetic field
$\vec{B}$ are possible: with the current {$\vec{\jmath}$} flowing
perpendicular to the magnetic field ($\vec{\jmath}\perp \vec{B}$) or
parallel with it ($\vec{\jmath}\parallel \vec{B}$). The Hall effect
plays a role in the case $\vec{\jmath}\perp \vec{B}$. In wide {\em
macroscopic} samples (superlattices) the standard method of the
description of the Hall effect, via the inversion of the conductivity
tensor, applies. This is no longer true for the double-well structures
the width of which is {\em microscopic}. This is why we present here a
novel theoretical approach to the calculation of the Hall effect,
based on its textbook definition: the Hall electric
field is caused by the accumulation of positive and negative charges
on the opposite sample edges, induced by the Lorentz force.

The difference between conductivities along and perpendicular to the
magnetic field direction was investigated experimentally in several
papers,~\,\cite{Simmons,Berk2,Ihn}. The results, obtained by solving
the Boltzmann equation~\,\cite{Raichev}, have been found in
a qualitative agreement with the experimental observation.  However, no
attention has been paid to the effect of the sample polarization due
to the transfer of electrons between wells -- to the Hall effect.

\begin{figure}[hbt]
\centering{\resizebox{7cm}{!}{\includegraphics{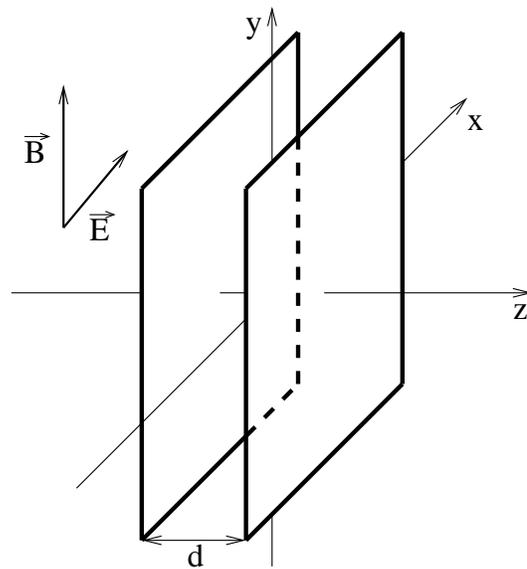}}}
\caption{Schematic picture of a double-layer system.  Directions of
applied electric and magnetic fields are indicated.}
\label{fig1}
\end{figure}

Let us consider the setup sketched in Fig.~\ref{fig1}. The electrons
are pushed by a Lorenz force from the left to the right well.
Resulting non-equilibrium charge distribution gives rise to a
potential that compensates for the Lorenz force in a similar way as
the Hall voltage does in macroscopic samples. In a limiting case of
very narrow wells, the double-well system can be viewed as a parallel
plate capacitor: the induced potential reduces to an inter-well Hall
voltage $U_H$, a potential difference between electron systems in
individual wells.

Note that in such an experimental arrangement the electron transport is
closely related to the chirality of diamagnetic currents flowing in
opposite directions in the left and right wells. In the thermodynamic
equilibrium, the two currents are exactly balanced giving a zero net
flow of electrons through the sample. In non-equilibrium states the
electron system is polarized, more electrons are in the right well and
an incomplete compensation of diamagnetic currents results in a
finite transport current.  The aim of this paper is to show that this
usually neglected correction to the in-plane electrical conductivity
gives a significant contribution for realistic double well structures.

For the sake of simplicity we employ the model of a double well
described in a paper~\,\cite{Hu}. Two coupled, strictly
two-dimensional electron layers confined in very narrow potential
wells are considered. Short-range scatterers, randomly distributed in
individual wells, are assumed to be responsible for a finite
relaxation time of electrons.

We use the {\em self-consistent} linear response theory (Kubo
formula) in the random phase approximation (RPA) (see
e.g.~\,\cite{RPA}) to obtain the non-equilibrium charge distribution,
the inter-layer Hall voltage and the conductivity tensor components.
\section{Model Hamiltonian}
In our model, the vector potential $\vec{A}=(zB,0,0)$ is used to
describe the influence of the in-plane magnetic field
$\vec{B}=(0,B,0)$ on the electron structure of a double-layer system.
The corresponding Hamiltonian reads
\begin{equation}
\hat{H}_0=\frac{1}{2m^*}\left(\hat{p}_x+|e|Bz\right)^2
         +\frac{1}{2m^*}\left(\hat{p}_y^2+\hat{p}_z^2\right)+V(z),
\label{hamilt}
\end{equation}
where $m^*$ denotes the electron effective mass and $V(z)$ is a
confining potential with two deep minima at $z=\pm d/2$, $V(d/2) =
V(-d/2)$. The subscript $0$ denotes that no scattering mechanism is
included. The form of (\ref{hamilt}) allows  separation of
variables, the $x,y$ dependent part of eigenstates of $\hat{H}_0$ is
the plain wave $|\vec{k}\rangle$ with the two-dimensional wave vector
$\vec{k}=(k_x,k_y)$.\footnote{In the rest of the paper all the
  vectorial notation corresponds to two-dimensional vectors lying in
  the layer planes. The only exception are the vectors $\vec{A}$ and
  $\vec{B}$ introduced above.}
To describe electron eigenstates depending on the
remaining third coordinate $z$ we restrict ourselves to the
tight-binding approximation employing only the lowest eigenstates
$|\varphi_{\alpha}(z)\rangle$ of uncoupled
wells~\,\cite{Raichev,Hu}. The index $\alpha = R,L$ distinguishes the
right and left well. Thus, a complete single-layer eigenstate reads
$|\varphi_{\alpha}, \vec{k}\rangle = |\varphi_{\alpha}(z)\rangle
\,|\vec{k}\rangle$ and
\begin{equation}
\hat{P}_{\alpha} =
\sum_{\vec{k}}|\varphi_{\alpha},\vec{k}\rangle
\langle \varphi_{\alpha},\vec{k}|
\end{equation} 
has a meaning of a projection operator to the states in the well
$\alpha$, $\hat{P}_L + \hat{P}_R = 1$. The Hamiltonian (\ref{hamilt})
is diagonal in the index $\vec{k}$ in our restricted basis and takes
the matrix form
\begin{equation}
\langle\vec{k}|\hat{H}_0|\vec{k}'\rangle=\delta_{\vec{k},\vec{k}'}
\pmatrix{
E_L(\vec{k}) &     t         \cr
    t         & E_R(\vec{k})    },
\label{hamiltMatrix}
\end{equation}
where $E_{L,R}(\vec{k})$ are single-well eigenenergies,
\begin{equation}
E_{L,R}(\vec{k}) = \frac{1}{2m^*}
\left(\hbar k_x\mp\frac{1}{2}|e|Bd\right)^2
+\frac{\hbar^2k_y^2}{2m^*}.
\label{eigene}
\end{equation}
The hopping integral $t$ is given by $t =
\langle\varphi_L|V(z)|\varphi_R\rangle =
\langle\varphi_R|V(z)|\varphi_L\rangle $.  Note that we neglected the
effect of the in-plane field on the energy spectra of individual
layers and that the $z$ coordinate has been replaced in (\ref{eigene}) by $\pm
d/2$ for the right and left layer, respectively.

The diagonalization of the matrix (\ref{hamiltMatrix}) yields a pair
of new eigenstates $|\chi^{(x)}_i(z)\rangle$. They depend only on the
$x$ component of $\vec{k}$ and the index $i$ equals $b$ for the bonding
state and $a$ for the antibonding state. The new basis $|\chi^{(x)}_i
\rangle$ is related to the basis of single-well eigenstates
$|\varphi_{\alpha}\rangle$ by
\begin{equation}
|\chi^{(x)}_{b,a}\rangle = 
\frac{1}{\sqrt{2}\,}\sqrt{1\pm\Delta}\;|\varphi_L\rangle
 \pm \frac{1}{\sqrt{2}\,}\sqrt{1\mp\Delta}\; |\varphi_R\rangle ,
\end{equation}
where $\Delta=\delta/\sqrt{\delta^2+t^2}$ and $\delta$ is the
abbreviation for $\hbar |e|dBk_x/2m^*$. Then the eigenenergies of
(\ref{hamilt})  reads
\begin{eqnarray}
E_{b,a}(\vec{k}) & = & E^{(x)}_{b,a}(k_x) + \frac{\hbar^2k_y^2}{2m^*} \; , \\
E^{(x)}_{b,a}(k_x) & = &  \frac{\hbar^2k_x^2}{2m^*} +
       \frac{e^2d^2B^2}{8m^*}\mp\sqrt{\delta^2+t^2}. 
\nonumber
\end{eqnarray}

The model is characterized by two parameters, the inter-layer distance
$d$ and the hopping integral $t$. It is a good approximation when the
cyclotron energy and the tunnel coupling are small in comparison with
the single-well quantization energies.

The Hamiltonian $\hat{H}$ of the system with short-range scatterers is
obtained by adding the impurity potential $\hat{V}_{imp}$ to
$\hat{H}_0$. As mentioned above, we assume that impurities are
distributed in both the left and right wells at random. To be more
specific, the scattering on an individual impurity is considered to be
intra-well (diagonal in the layer index $\alpha$) and isotropic in
$\vec{k}$ direction.  The finite lifetime and transport relaxation
time result from replacing the quantities characteristic for a given
configuration of scatterers by these quantities averaged over all
possible configurations.
\section{Charge Redistribution}
Let us first turn attention to the self-consistent procedure of
establishing the non-equilibrium charge distribution and the Hall
potential $U_H$ in the RPA, as this is the novel feature of our
approach to the theoretical description of magnetotransport in
microscopically narrow systems. The basic scheme of our derivation is
outlined below, more details are presented in the Appendix.

The standard Kubo formula describing the linear response
to a time-dependent perturbation operator $\hat{H}_p$
is employed, our formalism is close to that used in\,~\cite{Bastin}.
In our case, the operator $\hat{H}_p$ is given as a sum of two parts:
the externally applied potential represented by $\hat{H}^{(1)}_p$ and
the self-consistent electric field of non-equilibrium electrons
described by $\hat{H}^{(2)}_p$.

The external homogeneous electric field $\vec{\cal{E}} = ({\cal{E}}_x,
{\cal{E}}_y)$ is included into the vector potential $\vec{A}$. Then
the corresponding part of $\hat{H}_p$ takes the form
\begin{equation}
\hat{H}_p^{(1)}(t)=\frac{e}{i\omega}
\hat{\vec{v}}\cdot\vec{\cal{E}}
e^{i\omega t},
\label{pertX}
\end{equation}
where $\hat{\vec{v}} = (\hat{v}_{x}, \hat{v}_{y})$ is the velocity
operator and $\omega$ denotes the frequency.  This perturbation yields
the static polarization of the sample and induces the stationary
current in the limit $\omega \to 0$.

The second part of the perturbation Hamiltonian, denoted as
$\hat{H}_p^{(2)}$, describes the Hartree potential derived from the
non-equilibrium electron density by solving the Poisson equation. In
our model, this potential reduces to the potential difference $U_H$
between two 2D electron layers and the operator $\hat{H}_p^{(2)}$ can
be written as
\begin{equation}
\hat{H}_p^{(2)} = e \frac{U_H}{d}\, \hat{z}, \,\,\,\,
\hat{z} = \frac{d}{2}\left(\hat{P}_R - \hat{P}_L\right) .
\label{pertZ}
\end{equation}
Here $\hat{z}$ stands for an operator of the $z$ coordinate in the
representation of single-layer eigenstates $|\varphi_{\alpha}\rangle$
and $U_H$ has to be determined self-consistently.

We introduce $\delta Q_{\alpha}$, the non-equilibrium charge density
per unit area of a well $\alpha $, as an expectation value of the
operator
\begin{equation}
\label{chargeR}
\hat{Q}_{\alpha} = e\hat{P}_{\alpha} =
e\, \sum_{\vec{k}}|\varphi_{\alpha},\vec{k}\rangle
\langle \varphi_{\alpha},\vec{k}|\, ,
\end{equation} 
evaluated using the non-equilibrium density matrix. Due to the charge
conservation $\delta Q_R$ is equal to $ -\delta Q_L$, and we can write
$\delta Q$ instead of $\delta Q_R$ ($-\delta Q$ instead of $\delta
Q_L$), for convenience.  

In our model, the Hall potential calculated in the Hartree
approximation, $U_H$, is related to the local non-equilibrium charge
density $\delta Q$ by a particularly simple formula
\begin{equation}
 \delta Q  = \varepsilon \frac{U_H}{d}.
\label{charge}
\end{equation} 
This expression corresponds to the parallel plate capacitor in which
the potential difference between plates is fully determined by their
distance, the excess charge on one particular plate and the dielectric
constant $\varepsilon $ of the insulator between plates.

Since $\hat{H}_p$ is a sum of $\hat{H}_p^{(1)}$ and $\hat{H}_p^{(2)}$,
also the charge density $\delta Q_{\alpha}$ can be written as a sum of
two contributions $\delta Q_{\alpha}^{(1)}$ and $\delta
Q_{\alpha}^{(2)}$.  We get the response to $\hat{H}_p^{(1)}$ in the
form
\begin{eqnarray}
\label{chargeX}
\delta && Q_{\alpha}^{(1)}  =
i\hbar e^2 {\cal{E}}_x
\int dE f_{0}(E) \\
&& \times\mathop{\rm Tr}
\left \langle \delta(E-\hat{H}) \left [
\hat{P}_{\alpha} \frac{d \hat{G}^+(E)}{d E} \hat{v}_x -
\hat{v}_x \frac{d \hat{G}^-(E)}{d E} \hat{P}_{\alpha} \right ]
\right \rangle ,
\nonumber
\end{eqnarray}
in the limit $\omega\to 0$.  Here $\langle\cdots\rangle$ denotes the
configuration averaging and $f_0$ stands for the Fermi-Dirac
distribution function.  As expected, the applied electric field
${\cal{E}}_y$ does not induce any excess charge.

The response to the operator $\hat{H}_p^{(2)}$,  defined by
equation~\,(\ref{pertZ}), yields
\begin{eqnarray}
\label{chargeZ}
\delta && Q_{\alpha}^{(2)}   =
-\frac{e^2}{d}  U_H \int dE f_{0}(E)  \\
&& \times\mathop{\rm Tr} \left \langle \delta(E-\hat{H})
\left [ \hat{z} \hat{G}^-(E) \hat{P}_{\alpha} +
\hat{P}_{\alpha} \hat{G}^+(E) \hat{z}\right] \right \rangle .
\nonumber
\end{eqnarray}
The equations (\ref{chargeX}) and (\ref{chargeZ}) relate the induced
charge densities to the electric fields ${\cal{E}}_x$ and
$-U_H/d$ and their structure reminds the equation
(\ref{charge}). Therefore, if we consider them as definitions of the
generalized dielectric functions $\varepsilon^{(1)}$ and
$\varepsilon^{(2)}$, we can write
\begin{equation}
 \delta Q^{(1)} = \varepsilon^{(1)} {\cal{E}}_x ,\,\,\,\,
 \delta Q^{(2)} = \varepsilon^{(2)} \left(-\frac{U_H}{d}\right) 
\end{equation}
where again the layer index is omitted and we have in mind the partial
densities in the right well.
  
Making use of $\delta Q = \delta Q^{(1)} + \delta Q^{(2)}$, the
decomposition of the local charge density $\delta Q$ into two
components, the self-consistent equation (\ref{charge}) takes the form
\begin{equation}
  \varepsilon^{(1)} {\cal{E}}_x - \varepsilon^{(2)}\, \frac{U_H}{d} =
  \varepsilon \, \frac{U_H}{d},  
\end{equation}
and its solution reads
\begin{equation}
 \frac{U_H}{d} = \frac{\varepsilon^{(1)}}{\varepsilon^{(2)}
+\varepsilon}\, {\cal E}_x .
\end{equation}
This formula relates the self-consistent Hartree field of electrons to the
$x$\,~component of the externally applied electric field $\vec{\cal E}$,
as  we have  anticipated in the introduction.
 
\section{In-plane conductivity}
With the established relations between $U_H$, $\delta Q$ and ${\cal
E}_x$, the calculation of the conductivity is a straightforward
procedure. The components of the tensor $\tensor{\sigma}$ are
determined through the expectation values of the current operator
\begin{equation}
\label{current}
\hat{\vec{\jmath}} =  e\, \hat{\vec{v}} +
\frac{e^2 \vec{\cal E}}{i m^* \omega} e^{i \omega t}, 
\end{equation}
calculated using the non-equilibrium density matrix given by a
response to the perturbation $\hat{H}_p$ in the limit $\omega \to 0$.

Similarly as the generalized dielectric function $\varepsilon$, the
conductivity tensor can be written as a sum of two contributions
$\tensor{\sigma}^{(1)}$ and $\tensor{\sigma}^{(2)}$.  The response to
$\hat{H}_p^{(1)}$ leads to the standard expressions~\,\cite{Bastin}
for the static conductivity $\tensor{\sigma}^{(1)}$.  We can write for
its components
\begin{eqnarray}
\label{condXX0}
\sigma_{ss}^{(1)} = && \pi\hbar e^2 \int dE
\left ( - \frac{df_0(E)}{dE} \right ) \\
&& \quad \times \mathop{\rm Tr}
\left \langle \delta(E-\hat{H})
\hat{v}_{s}\delta(E-\hat{H})\hat{v}_{s}
\right \rangle \, , \nonumber
\end{eqnarray}
where $s$ stands for $x$ or $y$.


The response to the self-consistent field described by the
perturbation Hamiltonian $\hat{H}_p^{(2)}$ gives a non-zero value only
for the current flowing along the $x$ direction. The corresponding
contribution to the conductivity reads
\begin{eqnarray}
\label{deltaXX1}
&&  \sigma^{(2)}_{xx}= e^2
\frac{\varepsilon^{(1)}}{\varepsilon^{(2)}+\varepsilon}
\int dE  f_{0}(E) \\
&&\quad\times\mathop{\rm Tr}
\left \langle \delta(E-\hat{H}) \left [
\hat{z} \hat{G}^-(E)\hat{v}_x + \hat{v}_x\hat{G}^+(E) \hat{z}
\right]\right \rangle \; . \nonumber
\end{eqnarray}
Thus, while the conductivity $\sigma_{yy}$ is determined only by the
external field ${\cal E}_y$ and $\sigma_{yy} =\sigma_{yy}^{(1)}$, in
the case of the $x$ direction also the self-consistent Hall field
$U_H$ helps to conduct the current and $\sigma_{xx} =
\sigma_{xx}^{(1)} + \sigma_{xx}^{(2)}$.
\section{Short-range scatterers in Born approximation}
The primary aim of this paper is to estimate the importance of the
contribution of $\sigma^{(2)}_{xx}$ to the conductivity
$\sigma_{xx}$. 

We have performed numerical calculation of these quantities for the
realistic parameters of a bilayer system.  The notably difficult problem
is to model the electron scattering on impurities, as we have only
limited knowledge about the nature of scatterers and their
distribution in the sample. For simplicity, we assume the short-range
scattering of electrons on impurities randomly distributed in both the
left and right wells, as mentioned above.  The concentration of
impurities is assumed very low and the weak scattering on an
individual impurity is treated in the Born approximation. The strength
of the scattering on an individual impurity and the impurity
concentrations are taken as adjustable parameters. Only the
non-self-consistent version of the Born approximation is used which
leads to inaccurate description close to singularities in the field
dependence of $\tensor{\sigma}$.

In the course of calculations, we apply the standard procedure of
averaging the resolvents and their products over all possible
configurations of impurities. We assume $\langle
V_{imp}\rangle = 0$ as usual. The averaged resolvent
$\langle\hat{G}^{\pm}(E)\rangle$ is approximated by
$\overline{G}^{\pm}(E)$, which satisfies the coupled equations
\begin{equation}
\label{D}
\overline{G}^{\pm}(E) = \frac{1}{E - \hat{H}_0 - \hat{\Sigma}^{\pm}}\, ,
\end{equation}
\begin{equation}
\label{S}
\hat{\Sigma}^{\pm} =  \langle\hat{V}_{imp} \overline{G}^{\pm}(E)
\hat{V}_{imp}\rangle \, ,
\end{equation}
where $\hat{\Sigma}^{\pm}$ is the self-energy operator in the Born
approximation, i.e. determined to the second order in the perturbation
$\hat{V}_{imp}$.

The products $\langle\hat{G}^{\pm} \hat{v}_s \hat{G}^{\pm}\rangle$, $s =
x,y$, appear in the expressions for conductivity.  We approximate this
expressions by the vertex functions $\hat{K}_s^{\pm\pm}$, which must be
found by solving the Bethe-Salpeter equations. In the Born approximation,
the equations take the form
\begin{equation}
\label{K}
\hat{K}_s^{\pm\pm} = \overline{G}^{\pm}\left(\hat{v}_s +
\langle \hat{V}_{imp}\hat{K}_s^{\pm\pm}\hat{V}_{imp}\rangle\right) 
\overline{G}^{\pm}\, ,
\end{equation}
where $\langle \hat{V}_{imp}\hat{K}_s^{\pm\pm}\hat{V}_{imp}\rangle$
are the vertex corrections to the velocity component $\hat{v}_s$.

To proceed further, we accepted additional simplifications motivated
by the very low scattering rate in high-mobility samples.  
Namely, the real part of the self-energy $\hat{\Sigma}^{\pm}$ (a
correction to the eigenenergies) is neglected and only the terms of
the lowest order of the scattering rate are kept in all expressions.

In this approximation,
the imaginary part of the self-energy $\hat{\Sigma}^{\pm}$, denoted by
$\hat{\Gamma}$, is diagonal in the spectral representation of the
Hamiltonian $H_0$ and proportional to the densities of states,
$g_{\alpha} = m^*/(\pi\hbar^2)$, of the uncoupled electron layers,
\begin{eqnarray}
\Gamma_{b,a}(E,k_x) && =
\langle \chi_{b,a}^{(x)}|\hat{\Gamma}(E)|\chi_{b,a}^{(x)}\rangle  \\
&& = \pi \gamma_L (1 \pm \Delta) g_L +
\pi \gamma_R (1 \mp \Delta) g_R \, . \nonumber
\end{eqnarray}
The parameters $\gamma_{\alpha}$ represent the intra-well scattering
rate and are given by the square of matrix elements of the
single-impurity potential, multiplied by the concentrations of
impurities.  Thus, $\Gamma_{b,a}$ is determined as a weighted sum of
scattering inside the left and right subsystems with the weights $1
\pm \Delta$. 

It is also worth to know how the resulting quantities depend on the
scattering rate. While $\sigma^{(1)}_{ss}$ and
$\varepsilon^{(1)}$ would diverge in samples without impurities and
the leading term of their power expansion is of the order $1/\Gamma$,
$\sigma_{ss}^{(1)} \propto 1/\Gamma$ and $\varepsilon^{(1)}\propto
1/\Gamma$, the expressions obtained as the response to $U_H$ will stay
finite in this limit, $\varepsilon^{(2)}\propto 1$. Note that in spite
of it $\sigma_{xx}^{(2)} \propto 1/\Gamma$ as $\sigma_{xx}^{(2)}$ is
related to $\varepsilon^{(1)}$ through equation\,(\ref{deltaXX1}).
 
The vertex corrections vanish for the velocity component $v_y$, i.e.
the product $\langle\hat{G}^{\pm}\hat{v}_y\hat{G}^{\pm}\rangle$ can be
approximated by $\overline{G}^{\pm} \hat{v}_y \overline{G}^{\pm}$.
The matrix elements of $\hat{v}_y$ are symmetric in $\vec{k}$ space,
$\langle\alpha|\hat{v}_y(-\vec{k})|\alpha\rangle =
-\langle\alpha|\hat{v}_y(\vec{k})|\alpha\rangle$, their symmetry and
the isotropy of the scattering by randomly distributed short-range
scatterers are responsible for vanishing the vertex corrections in
this case.  

This is not true for
$\langle\hat{G}^{\pm}\hat{v}_x\hat{G}^{\pm}\rangle$.  Due to the
breaking of the time reversal symmetry by $\vec{B}\perp x$,
$\langle\alpha|\hat{v}_x(-\vec{k})|\alpha\rangle \not =
-\langle\alpha|\hat{v}_x(\vec{k})|\alpha\rangle$ and the vertex
corrections have to be taken into account. We must solve two coupled
Bethe-Salpeter equations for the vertex functions.  Unfortunately, for
our very simple model they are linearly dependent. Similar
complication was found also in~\,\cite{Raichev}, where the same model
was used to describe the electron transport by the Boltzmann
equation. The second independent equation was supplied by the charge
conservation rule $\delta Q_R = -\delta Q_L$.
%
\section{Results and discussion}
The parameters corresponding to the sample B studied experimentally by
Simmons et al.~\,\cite{Simmons} are used in our model calculations.
The distance between wells $d$ equals $135\,${\AA} and the hopping
integral $t$ is supposed to be  $0.9\,$meV. The density of electrons $N_e =
2.4\times 10^{11}\,$cm${}^{-2}$ yields the Fermi energy $4.29\,$meV.

The applied in-plane magnetic field qualitatively changes  the
topology of Fermi contours~\,\cite{Simmons,Raichev}.  At zero field,
the Fermi contours are two concentric circles. The larger circle
corresponds to bonding states, the smaller one to antibonding
states. The applied in-plane magnetic field shifts the centers of
circles in the opposite directions in $k$ space and gradually changes
their shapes. The Fermi contour of the bonding subband acquires the
shape of a ``peanut'', the Fermi line of the antibonding subband evolves
into a ``lens'' shape. The area of the lens is reduced by the
increasing in-plane field.  At the critical field $B_{c,1} = 7.4\,$T
the lens vanishes and the transition from two-component to
one-component system occurs.  The peanut splits at the higher critical
field $B_{c,2} = 9.3\,$T into two parts, the left and right electron
layers become decoupled and two disconnected ovals represent the Fermi
contours of individual layers above that field. The changes in the
connectivity of Fermi lines are reflected in the field dependence of
the density of electronic states (DOS) as van Hove singularities at
the critical fields.  
\begin{figure}[htb]
\resizebox{8cm}{!}{\rotatebox{-90}{\includegraphics{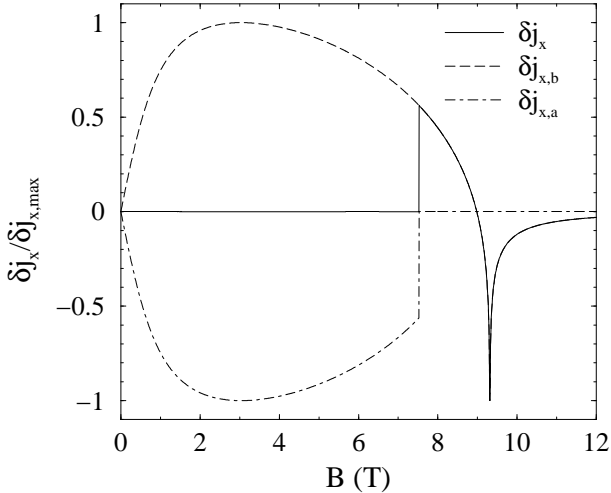}}}
\caption{Calculated magnetic field dependence of the equilibrium
diamagnetic current of electrons on the Fermi contour flowing in the
right well. Except of the total current also the contributions from
the bonding and antibonding subbands are shown: $\delta j_x =
\delta j_{x,b} + \delta j_{x,a}$.}
\label{fig2}
\end{figure}

As we have mentioned in the introduction, the incomplete balance of
chiral diamagnetic currents flowing in the right and left wells is
behind the origin of the electron transport in the $x$ direction. In
the linear response theory we implicitly assume that applied external
fields are infinitesimally weak.  Therefore, also the polarization is
weak and only the electrons with energies in a narrow interval around
the Fermi energy contribute to the difference between the currents in
the left and right wells.  For this reason, Fig.~\ref{fig2} presents
the calculated equilibrium diamagnetic current $\delta j_{R,x}$
flowing in the right well only for electrons on the Fermi contour.
(The current of the opposite direction flows in the left well and
$\delta j_{L,x} = -\delta j_{R,x}$.) Together with the total current
$\delta j_x (\equiv \delta j_{R,x})$ also the contributions $\delta
j_{x,b}$ and $\delta j_{x,a}$ from the bonding and antibonding
subbands are shown. In our simplified model $\delta j_{x,b}$ and
$\delta j_{x,a}$ exactly cancels out for magnetic fields less than
$B_{c,1}$.  Above $B_{c,1}$ the current $\delta j_x \equiv \delta
j_{x,b}$ decreases and changes sign before the magnetic field reaches
$B_{c,2}$. At that field $\delta j_x$ diverges and for $B>B_{c,2}$
returns slowly back to zero value expected for the decoupled
layers. The sharp minima and maxima of the total $\delta j_x$ reflect
qualitative changes of the topology of Fermi contours, caused by the
applied in-plane magnetic field, similarly as a field dependence of
the DOS.

\begin{figure}[htb]
\resizebox{8cm}{!}{\rotatebox{-90}{\includegraphics{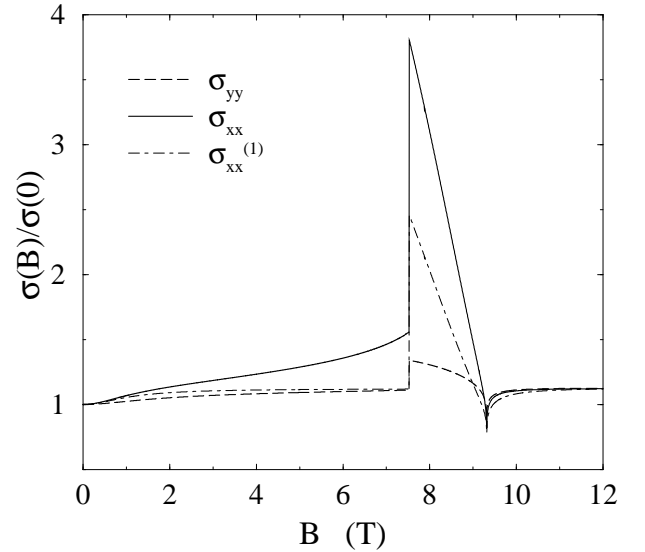}}}
\caption{Magnetic field dependences of the conductivity 
components $\sigma_{xx}$ and $\sigma_{yy}$. $\sigma^{(1)}_{xx}$
is the conductivity calculated without the Hall correction 
($\sigma^{(2)}_{xx}$ neglected).}
\label{fig3}
\end{figure}

The results for the conductivity tensor components $\sigma_{xx}$ and
$\sigma_{yy}$, calculated assuming zero temperature, are presented in
Fig.~\ref{fig3}. The overall similarity of curves in Fig.~\ref{fig3}
to $\delta j_x$ is striking. The van Hove singularities at the
critical fields $B_{c,1}$ and $B_{c,2}$ seen on $\delta j_x$ appear
also on $\sigma_{xx}$ and $\sigma_{yy}$. Moreover, both components of
the conductivity are small below $B_{c,1}$ where $\delta j_x$
vanishes, are decreasing functions for $B$ between $B_{c,1}$ and
$B_{c,2}$, and increase above $B_{c,2}$ similarly as $\delta j_x$.  But
there are, of course, many differences which stem from the more
detailed description of the scattering process and response to the
external fields, as presented in previous paragraphs.

The dependence of the conductivity components $\sigma^{(1)}_{xx}$ and
$\sigma_{yy}$ on the magnetic field coincides with that already
obtained by solving the Boltzmann equation~\,\cite{Raichev}. In this
case the origin of the van Hove singularities is related rather to DOS
than to $\delta j_x$. The difference between $\sigma^{(1)}_{xx}$ and
$\sigma_{yy}$ originates in the Fermi contours anisotropy and the
related anisotropy of the effective mass and the relaxation time.  The
novel ``Hall'' correction $\sigma^{(2)}_{xx}$ to the conductivity
$\sigma_{xx}$ does not change the field dependence qualitatively, but
strengthens the difference between $\sigma_{xx}$ and $\sigma_{yy}$
substantially.
 
\begin{figure}[htb]
\resizebox{8cm}{!}{\rotatebox{-90}{\includegraphics{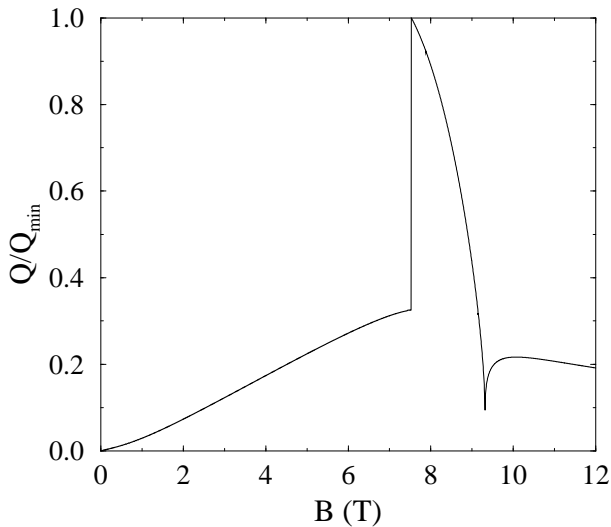}}}
\caption{Charging of the right layer when external electric field is
oriented in the $x$ direction.  Note that $Q$ is negative.}  
\label{fig4}
\end{figure}

The question arises whether one can distinguish between
$\sigma^{(1)}_{xx}$ and $\sigma^{(2)}_{xx}$ experimentally.  The
possibility can be to study the inter-layer charge redistribution by
measuring the change of the capacity between the bilayer system and
the gate electrode on the top of the sample. It should be proportional
to the current flowing through the sample perpendicularly to the
magnetic field.  The excess electric charge cumulated in the right
layer, induced by the applied electric field in the $x$-direction, is
shown in Fig.~\ref{fig4}. The maximum effect is expected when only a
bonding subband is occupied and the Fermi line has a
single-connected-peanut shape.
\noindent
To conclude: we have shown that the charge redistribution within a
double-layer system gives rise to the Hall-like contribution to the
conductivity which is so strong that it cannot be omitted. The
experimental verification of this Hall effect is however
nontrivial. Any potential contacts destroy the double-layer structure
heavily and a direct measurement of the potential difference between
layers can hardly be realized.  We believe that the solution is a
non-invasive measurement of the charge redistribution from the
capacity changes of gated structures.
\acknowledgments
This work has been supported by the Grant Agency of the Czech
Republic under Grant No. 202/01/0754.
\appendix
\section*{}
A Kubo formula evaluates the linear response of the density matrix
$\hat{f}(t)$ to a perturbation of the form
$\hat{H}_p(t)=-\hat{A}F(t)$, where $\hat{A}$ is a time-independent
operator and $F(t)$ is a function of time. The non-equilibrium matrix
deviates from the Fermi-Dirac distribution function $f_0(\hat{H})$ by
$\delta\hat{f}(t)$. The expectation value of an operator $\hat{B}$,
$\delta B(t) = \mathop{\rm Tr}\left(\hat{B}\delta\hat{f}(t)\right)$,
is related to the perturbation by
\begin{eqnarray}
\label{a1}
\delta B(t)=&&
 \frac{1}{i\hbar}\int_{-\infty}^0 dt' e^{\epsilon t'}\\
&&\times\mathop{\rm Tr}\left(f_0(\hat{H})
 \left[e^{\frac{i}{\hbar}\hat{H}t'}\hat{A}e^{-\frac{i}{\hbar}\hat{H}t'},
 \hat{B}\right]\right)F(t+t').\nonumber
\end{eqnarray}
Taking into account $\hat{H}_p^{(1)}$, given by (\ref{pertX}), we get 
\begin{eqnarray}
\label{a2}
\delta B(t)=&&\frac{eE_x}{\hbar\omega}e^{i\omega t}\int_{-\infty}^0 dt' 
 e^{\epsilon t'}
\left[\left(e^{i\omega t'}-1\right)+1\right]\\
&&\qquad\qquad\times\mathop{\rm Tr}\left(f_0(\hat{H})
 \left[e^{\frac{i}{\hbar}\hat{H}t'}\hat{v}_x e^{-\frac{i}{\hbar}\hat{H}t'},
 \hat{B}\right]\right)\nonumber,
\end{eqnarray}
where 1 was subtracted and added in square  brackets of the first
line of the expression.   

The integration by parts of the second term, corresponding to the
added 1 in an above sum, converts (\ref{a2}) to a form
\begin{eqnarray}
\label{dBperpartes}
\delta B(t)=
&&\frac{eE_x}{\hbar\omega}e^{i\omega t}\int_{-\infty}^0 dt'
 e^{\epsilon t'}\left(e^{i\omega t'}-1\right)\\
&&\qquad\times\mathop{\rm Tr}\left(f_0(\hat{H})
 \left[e^{\frac{i}{\hbar}\hat{H}t'}\hat{v}_x e^{-\frac{i}{\hbar}\hat{H}t'},
 \hat{B}\right]\right)\nonumber\\
&&+\frac{eE_x}{\hbar\omega} e^{i\omega t}
 \mathop{\rm Tr}\left(f_0
 (\hat{H})\left[\hat{x},\hat{B}\right]\right)\nonumber.
\end{eqnarray}
To derive (\ref{dBperpartes}), we employed $\hat{v}_x =
[\hat{x},\hat{H}]/i\hbar$ and the identity
\begin{equation}
\int_{-\infty}^0 dt' \epsilon e^{\epsilon t'}
 \mathop{\rm Tr}\left(f_0(\hat{H})
 \left[e^{\frac{i}{\hbar}\hat{H}t'}\hat{x} e^{-\frac{i}{\hbar}\hat{H}t'},
 \hat{B}\right]\right)=0,
\end{equation}
which can be proved replacing functions of Hamiltonian standing inside
the trace according to prescription
\begin{equation}
Z(\hat{H})=\int dE Z(E)\delta(E-\hat{H}),
\end{equation}
where $Z$ denotes such a function, and performing the
integration over the time variable. The result
\begin{eqnarray}
\int dE&&\int dE'
 \frac{i\hbar\epsilon}{E'-E+i\hbar\epsilon}\\
&&\times\Bigl[f_0(E)-f_0(E')\Bigr]
 \mathop{\rm Tr}\left[\delta(E-\hat{H})\hat{x}
              \delta(E'-\hat{H})\hat{B}\right]\nonumber
\end{eqnarray}
is an expression which  goes to zero when  $\epsilon \to 0$. 

A similar approach is applied to the first part of (\ref{a2}),
corresponding to the subtracted 1.  After the time integration we
introduce resolvents by $\hat{G}^{\pm}(E)=\left(E-\hat{H}\pm
i\hbar\epsilon\right)^{-1}$, and, letting $\omega\to 0$ to get a static
response, we obtain the Kubo formula in the form
\begin{eqnarray}
\label{last}
\delta B(t)=&&i\hbar eE_x\int dE f_0(E)\\
&&
 \times\mathop{\rm Tr}\left\{\delta(E-\hat{H})\left[
 \hat{B}\frac{d\hat{G}^+(E)}{dE}\hat{v}_x
 -\hat{v}_x\frac{d\hat{G}^-(E)}{dE}\hat{B}
 \right]\right\}\nonumber\\
&&+\lim_{\omega\rightarrow 0}\frac{eE_x}{\hbar\omega}e^{i\omega t}
 \mathop{\rm Tr}\left(f_0(\hat{H})\left[\hat{x},\hat{B}\right]\right).
 \nonumber
\end{eqnarray}

The expression (\ref{chargeX}) for $\delta Q_{\alpha}$ is derived by
introducing $\hat{Q}_{\alpha}$,  defined by equation (\ref{chargeR}),
into (\ref{last}) instead of $\hat{B}$, and taking into account that
$[\hat{x},\hat{Q}_{\alpha}] = 0$.  The conductivity formula
(\ref{condXX0}) is obtained when $\hat{B}$ is replaced by the current
operator (\ref{current}). In this case the last term of (\ref{last})
gives a non-zero contribution, which exactly compensates  the
equilibrium gauge current.

The expressions (\ref{chargeZ}) and (\ref{deltaXX1}) are derived in an
analogous way, taking into account the perturbation
Hamiltonian $\hat{H}_p^{(2)}$ instead of $\hat{H}_p^{(1)}$. In fact,
the derivation is simpler in this case.

\end{document}